\documentclass[aps,pre,twocolumn]{revtex4}
\usepackage[dvips]{graphics}
\usepackage{graphicx}
\usepackage{amsfonts}
\usepackage{amssymb}
\usepackage{amsmath}

\begin{document}

\title{Discrete Solitons and Vortices on Anisotropic Lattices}
\author{P.G. Kevrekidis$^1$, D.J.\ Frantzeskakis$^{2}$, R.\
Carretero-Gonz\'{a}lez$^3$, B.A. Malomed$^3$ and A.R. Bishop$^5$}
\affiliation{$^1$ Department of Mathematics and Statistics, University of Massachusetts,
Amherst MA 01003-4515, USA \\
$^{2}$ Department of Physics, University of Athens, Panepistimiopolis,
Zografos, Athens 15784, Greece \\
$^3$ Nonlinear Dynamical Systems Group\footnote{URL: {\tt http://nlds.sdsu.edu/}},
Department of Mathematics and Statistics,
and Computational Science Research Center\footnote{URL: {\tt http://www.csrc.sdsu.edu/}},
San Diego State University, San Diego CA, 92182-7720, USA \\
$^{4}$ Department of Interdisciplinary Studies, School of Electrical
Engineering, Faculty of Engineering, Tel Aviv University, Tel Aviv 69978,
Israel \\
$^5$ Theoretical Division and Center for Nonlinear Studies, Los Alamos
National Laboratory, Los Alamos, New Mexico 87545, USA}

\begin{abstract}
We consider effects of anisotropy on solitons of various types in
two-dimensional nonlinear lattices, using the discrete nonlinear
Schr{\"{o}}dinger equation as a paradigm model. For fundamental
solitons, we develop a variational approximation, which predicts
that broad quasi-continuum solitons are unstable, while their
strongly anisotropic counterparts are stable. By means of
numerical methods, it is found that, in the general case, the
fundamental solitons and simplest on-site-centered vortex
solitons (``vortex crosses") feature enhanced or reduced stability
areas, depending on the strength of the anisotropy. More
surprising is the effect of anisotropy on the so-called
``super-symmetric" intersite-centered vortices (``vortex
squares"), with the topological charge $S$ equal to the square's
size $M$:
we predict in an analytical form by
means of the Lyapunov-Schmidt theory, and confirm by numerical
results, that arbitrarily weak anisotropy results in dramatic
changes in the stability and dynamics in comparison with the
\emph{degenerate}, in this case, isotropic limit.
\end{abstract}

\date{Submitted to {\em Phys.\ Rev.\ E}, July 2005}
\maketitle

\section{Introduction and the model}

In the last two decades, nonlinear lattice (spatially discrete) systems have
been a very rapidly growing area of interest for a variety of applications
\cite{reviews}. Such systems arise in physical contexts encompassing,
\textit{inter alia}, beam dynamics in coupled waveguide arrays in nonlinear
optics \cite{reviews1}, the time evolution of fragmented Bose-Einstein
condensates (BECs) trapped in optical lattices (OLs) \cite{reviews2},
coupled cantilever systems in nano-mechanics \cite{sievers}, denaturation of
the DNA double strand in biophysics \cite{reviews3} and even stellar
dynamics in astrophysics \cite{voglis}.

One of the main objectives of the research in this field is to
achieve an understanding of intrinsically
localized states (discrete solitons). In
two-dimensional (2D) lattices, these are fundamental discrete
solitons \cite{solit} and discrete vortices (i.e., localized
states with an embedded nonzero phase circulation over a closed
lattice contour) \cite{vort1,vort2}. Most recently, a substantial
effort was dedicated to the experimental creation of both these
entities in photonic lattices induced in photorefractive crystals
(although these systems are only quasi-discrete). In
particular, fundamental and dipole solitons, soliton trains and
necklaces, and vector solitons have been reported \cite{esolit},
as well as vortex solitons \cite{evort}. Parallel developments in
the experimental studies of soliton patterns in BECs have also
been very substantial, leading to the creation of quasi-1D dark
\cite{becd}, bright \cite{becb} and gap \cite{becg} solitons. The
generation of 2D BEC solitons in OLs has been theoretically
demonstrated \cite{boris} to be feasible with the currently
available experimental technology \cite{old2d}.

A paradigm dynamical lattice model that appears in the
above-mentioned physical problems is the discrete nonlinear
Schr{\"{o}}dinger (DNLS) equation. Various applications of the
DNLS equation are well documented
\cite{reviews,reviews1,reviews2}. Besides being a generic
asymptotic form of a whole class of lattice models (for
small-amplitude nonlinear excitations), it finds direct
applications (where it furnishes extremely accurate description of
the underlying physics) in terms of arrayed (1D) or bunched (2D)
nonlinear optical waveguides, BECs trapped in strong OLs, and
crystals built of optical or exciton microcavities.

An interesting issue in this framework that has not received
sufficient attention concerns the influence of anisotropy on the
soliton dynamics in 2D lattices. Some of the settings mentioned
above are \textit{inherently} anisotropic, e.g., photorefractive
crystals \cite{solit,esolit}, while others (in particular, the
fragmented BECs trapped in strong OLs \cite{reviews2}) can be
easily rendered anisotropic by slight variations of control
parameters, such as intensities of laser beams that create two
sublattices which together constitute the 2D optical lattice.

The aim of this paper is to understand how the lattice anisotropy
affects 2D discrete solitons in the DNLS equation. Some findings
reported below are surprising, demonstrating that anisotropy
effects are not straightforward. The
straightforward expectation might be that weak anisotropy is a small
perturbation that possibly alters details of parametric
dependences of the observed phenomenology but does not change it
``structurally" (i.e., essentially the same dynamical features as
in the isotropic lattice occur, but at different positions in the
parameter space). We find that for the simplest soliton and vortex
structures this is indeed the case, while for more sophisticated
ones it is not. More specifically, we find that for especially
symmetric (so-called ``supersymmetric") vortices, with their center
set at an intersite position, and the topological charge equal to
the size of the vortex square frame (see below for details),
the isotropic lattice is a
\emph{degenerate} one, therefore even very weak anisotropy
fundamentally alters the stability and dynamical properties of
such structures. On the other hand, despite the delicate
organization of the supersymmetric vortices, they constitute a
structurally stable, i.e., physically meaningful, class of
objects.

We take the DNLS equation in the following form:
\begin{equation}
i\dot{u}_{n,m}=-\epsilon \Delta _{\alpha }u_{n,m}-|u_{n,m}|^{2}u_{n,m},
\label{deq1}
\end{equation}where $u_{n,m}(t)$ is the complex, 2D lattice field (the overdot stands for
its time derivative), $\epsilon $ is the lattice coupling constant, and
\begin{equation}  \label{deq2}
\begin{array}{rl}
\Delta _{\alpha }u_{n,m}=&\alpha \left( u_{n+1,m}+u_{n-1,m}\right)+u_{n,m+1}\\[2.0ex]
&+u_{n,m-1}-2(1+\alpha )u_{n,m},
\end{array}
\end{equation}
is the \emph{anisotropic} discrete Laplacian, which becomes
isotropic with $\alpha =1$. Note that, unlike the continuum limit, no scaling transformation
can cast the anisotropic DNLS equation into the isotropic form.
Equation (\ref{deq1}) conserves two dynamical invariants: the
Hamiltonian,
\begin{equation}
\begin{array}{rcl}
H &=&\sum_{n,m}\left[ \left(
u_{n,m+1}^{\ast }u_{n,m}+u_{n,m+1}u_{n,m}^{\ast }\right)
\right.\\[2.0ex]
&& \left. +\alpha
\left( u_{n+1,m}^{\ast }u_{n,m}+u_{n+1,m}u_{n,m}^{\ast
}\right) \right. \\[2.0ex]
&&\displaystyle
-\left. \left( \frac{\Lambda }{\epsilon }-2-2\alpha \right) \left\vert
u_{n,m}\right\vert ^{2}+\frac{1}{2\epsilon }\left\vert u_{n,m}\right\vert
^{4}\right] ,
\end{array}
\end{equation}
and norm,
\begin{equation}
N=\sum_{n,m}\left\vert u_{n,m}\right\vert ^{2},  \label{N}
\end{equation}
where $\Lambda$ is the frequency of the internal mode (equivalently
the chemical potential in the context of BECs or the propagation
constant in the context of optical waveguide arrays).

Stationary solutions to Eq.\ (\ref{deq1}) will be sought
as
\begin{equation}
u_{n,m}=u_{n,m}^{(0)}\exp (i\Lambda t),  \label{stationary}
\end{equation}
which leads to a stationary finite-difference equation,
\begin{equation}
\Lambda u_{n,m}^{(0)}=\epsilon \Delta _{\alpha }u_{n,m}^{(0)}-\left\vert
u_{n,m}^{(0)}\right\vert ^{2}u_{n,m}^{(0)}  \label{stateqn}
\end{equation}
(generally speaking, the discrete functions $u_{n,m}^{(0)}$ may be complex).
In the case of fundamental-soliton solutions, we will apply the
variational approximation (VA) to the \emph{real} version of Eq.\ (\ref{stateqn}),
which is based on the fact that it can be derived
from the Lagrangian,
\begin{equation} \label{L}
\begin{array}{rl}
L=&\displaystyle
\sum_{n,m}\left[u_{n,m+1}u_{n,m}+\alpha u_{n+1,m}u_{n,m}
-\phantom{\frac{A}{A}}\right. \\[4.0ex]
&\displaystyle\left.\left( \frac{\Lambda
}{2\epsilon }-1-\alpha \right) u_{n,m}^{2}+\frac{1}{4\epsilon
}u_{n,m}^{4}\right] .
\end{array}
\end{equation}

After analyzing fundamental solitons by means of the VA, we will
construct discrete solitons in the anisotropic model and will
study their stability by means of numerical methods. For the
numerical procedure, our starting point is always the
anti-continuum (AC) limit corresponding to $\epsilon =0$
\cite{mackay}, where configurations of interest can be constructed
at will as appropriate combinations of on-site states, which are
either $u_{n,m}=\sqrt{\Lambda }\exp (i\Lambda t)$ with $\Lambda
>0$ at excited sites, and $u_{n,m}\equiv 0$ at non-excited ones,
cf.\ Eqs.\ (\ref{stationary}) and (\ref{stateqn}) for the general
case, $\epsilon >0$. The stability of the solitons is then
analyzed by linearizing Eq.\ (\ref{deq1}) for perturbations around
a stationary solution $u_{n,m}^{(0)}e^{i\Lambda t}$,
\begin{equation}
u_{n,m}=\left[ u_{n,m}^{(0)}+\delta \cdot \left( a_{n,m}e^{\lambda
t}+b_{n,m}e^{\lambda ^{\ast }t}\right) \right] e^{i\Lambda t},
\label{lambda}
\end{equation}
where $\delta $ is an infinitesimal perturbation amplitude of the
perturbation, and $\lambda $ is its eigenvalue. The Hamiltonian
nature of the system dictates that if $\lambda $ is an eigenvalue,
then so also are $-\lambda $, $\lambda ^{\ast }$ and $-\lambda
^{\ast }$ (in the stable case, $\lambda $ is imaginary, hence this
symmetry yields only two different eigenvalues, $\lambda $ and
$-\lambda $). Clearly, the stationary solution is unstable if
at least one pair of eigenvalues features nonvanishing real parts.

It is noteworthy that the instability against perturbations
corresponding to purely real eigenvalues $\lambda $ in Eq.\ (\ref{lambda})
can be predicted by the \textit{Vakhitov-Kolokolov}
(VK) criterion \cite{VK}: a soliton family, characterized by the
dependence $N(\Lambda )$ [recall $N$ is the solution's norm
defined by Eq.\ (\ref{N})], may be stable under the condition
$dN/d\Lambda >0$, and is definitely unstable in the opposite case.
In particular, this criterion (as well as the VA)
was found to be very useful and
quite reliable  in the investigation of 2D
solitons in the Gross-Pitaevskii equation for BECs in 2D and
quasi-1D periodic OL potentials \cite{Bakhtiyor}, and even in 2D
quasi-periodic potentials (such as the Penrose tiling among others)
\cite{Bakhtiyor2}.

Our study of different states in the anisotropic model and their
properties is structured as follows. In Section II, we present the
VA for fundamental solitons. In Section III, discrete solitons and
vortex crosses with the topological charge $S=1$ are considered,
which are only perturbatively (weakly) affected by the anisotropy.
In the following two sections, we will define and consider
special ``super-symmetric" configurations, with $S=1$ and $S=2$,
respectively, and compare them with simpler cases. Finally, in
section VI we summarize findings and present our
conclusions.

\section{Variational approximation for fundamental solitons}

As was shown in Ref.\ \cite{MIW} for the one-dimensional DNLS
equation (see also Ref.\ \cite{MIW2}, for a more rigorous variational approach
applied to higher dimensional solitons in the isotropic case), the
only analytically tractable variational \textit{ansatz} for
stationary fundamental solitons may be based on the following
cusp-shaped expression (in the 2D case, it has the shape of a
cross cusp),\begin{equation} u_{n,m}^{(0)}=A\exp \left(
-a|n|-b|m|\right) ,  \label{ansatz}
\end{equation}with positive parameters $a$ and $b$, that determine the widths of the
soliton in the horizontal and vertical directions, and an
arbitrary amplitude $A$. Note that expression (\ref{ansatz}) is
indeed an \emph{exact solution} to the linearized version of Eq.\ (\ref{stateqn}),
which describes soliton tails, if $\Lambda $ is
linked to $a$ and $b$ by the dispersion relation,\begin{equation}
\Lambda =2\epsilon \left[ \alpha \sinh ^{2}(a/2)+\sinh
^{2}(b/2)\right] . \label{disp}
\end{equation}

The substitution of ansatz (\ref{ansatz}) makes it possible to
calculate the corresponding effective Lagrangian explicitly. First
of all, it is convenient to eliminate the amplitude in favor of
the norm (\ref{N}). Indeed, the substitution of the ansatz in the
definition of $N$ yields $A^{2}=N\tanh a\cdot \tanh b$. After
this, the effective Lagrangian becomes
\begin{eqnarray}
L_{\mathrm{eff}} &=&N~\left( \alpha \mathrm{sech}a+\mathrm{sech}b\right)
-\left( \frac{\Lambda }{2\epsilon }+1+\alpha \right) N  \notag \\
&+&\frac{N^{2}}{16\epsilon }\frac{\cosh \left( 2a\right) \cosh \left(
2b\right) \sinh a \sinh b}{\cosh ^{3}(a) \cosh ^{3}(b)}~.
\label{Leff}
\end{eqnarray}
Variational equations for the stationary profile are obtained from here
in the form
\begin{equation}
\frac{\partial L_{\mathrm{eff}}}{\partial N}=\frac{\partial
L_{\mathrm{eff}}}{\partial a}=\frac{\partial
L_{\mathrm{eff}}}{\partial b}=0. \label{general}
\end{equation}
In the general case, the explicit form of these equations is quite
cumbersome (this will be treated numerically, see below).
A detailed analysis is possible in two special cases, as
specified below.

First is the case of small $a$ and $b$ ($a,b\ll 1$), which implies broad
solitons. Then, the expansion of the effective Lagrangian (\ref{Leff}) yields
\begin{eqnarray}
L_{\mathrm{eff}} &\approx &-\frac{\Lambda }{2\epsilon }N+\frac{1}{2}N\left(
-b^{2}-\alpha a^{2}+\frac{5}{12}b^{4}+\frac{5\alpha }{12}a^{4}\right)  \notag
\\
&&+\frac{N^{4}}{16\epsilon }\left(
ab+\frac{2}{3}a^{3}b+\frac{2}{3}ab^{3}\right) ,  \label{expansion}
\end{eqnarray}and the variational equations (\ref{general}) following
from Eq.\ (\ref{expansion}) generate the following solution:
\begin{eqnarray}
N &=&16\epsilon \sqrt{\alpha }\left( 1-\frac{7}{8}\frac{\alpha +1}{\alpha }
\frac{\Lambda }{\epsilon }\right) ,  \label{NLambda} \\
a^{2} &=&\frac{\Lambda }{2\epsilon \alpha },~b=\sqrt{\alpha }a.  \label{ab}
\end{eqnarray}
As follows from these expressions, the underlying assumptions $a,b\ll 1$
indeed hold (i.e., the approximation is self-consistent) under the condition
\begin{equation}
\Lambda \ll \left\{
\begin{array}{cc}
\alpha \epsilon , & \mathrm{if~}\alpha ~\,_{\sim }^{>}~1, \\
\epsilon , & \mathrm{if~}\alpha \gg ~1.\end{array}\right.
\label{condLambda}
\end{equation}
The broad (quasi-continuum) solitons predicted in this approximation are
\emph{unstable} according to the VK criterion, as Eq.\ (\ref{NLambda})
immediately shows that $dN/d\Lambda <0$.

Note that the expansion of the dispersion relation (\ref{disp}) for the same
case of small $a$ and $b$ yields $\alpha a^{2}+b^{2}=\Lambda /\epsilon $. It
is noteworthy that this relation, although derived independently of the
variational equations, is consistent with Eq.\ (\ref{ab}).

Another tractable case is that of a \emph{strongly anisotropic}
soliton, which is broad (quasi-continuum) in either direction and
narrow in the other, i.e., it corresponds to $a\ll 1,b\gg 1$, or
vice versa. If $a$ is small and $b$ is large, the variational
equations (\ref{general}) yield the following
results:
\begin{equation} a=\sqrt{\frac{\Lambda }{3\alpha \epsilon
}},~\sinh \left( \frac{b}{2}\right) =\sqrt{\frac{\Lambda
}{\epsilon }},~N^{2}=\frac{4}{3}\epsilon \alpha \Lambda .
\label{aniso}
\end{equation}
These results are consistent with the underlying assumptions
($a\ll 1,b\gg 1$) under the conditions
\begin{equation}
1\ll \Lambda /\epsilon \ll \alpha .  \label{cond}
\end{equation}
On the contrary to the broad solitons given above by Eqs.\ (\ref{NLambda})
and (\ref{ab}), Eqs.\ (\ref{aniso}) show that the anisotropic
solitons are \emph{stable} as per the VK criterion, as they
obviously meet the condition $dN/d\Lambda >0$.

For the opposite strongly anisotropic case, with $a\gg 1$ and
$b\ll 1$, the result is\begin{equation} b=\sqrt{\frac{\Lambda
}{3\epsilon }},~\sinh \left( \frac{a}{2}\right)
=\sqrt{\frac{\Lambda }{\alpha \epsilon
}},~N^{2}=\frac{4}{3}\epsilon \Lambda , \label{aniso2}
\end{equation}cf. Eqs.\ (\ref{aniso}). These expressions comply with the underlying
assumptions $a\gg 1,b\ll 1$ provided that
\begin{equation}
\alpha \ll \Lambda /\epsilon \ll 1,  \label{cond2}
\end{equation}cf. Eq.\ (\ref{cond}). Similarly to the solution
of Eq. (\ref{aniso}),
the one of Eq. (\ref{aniso2}) obviously meets the VK stability criterion.

\begin{figure}[t]
\centerline{
\includegraphics[width=6.cm,height=4cm,angle=0,clip]{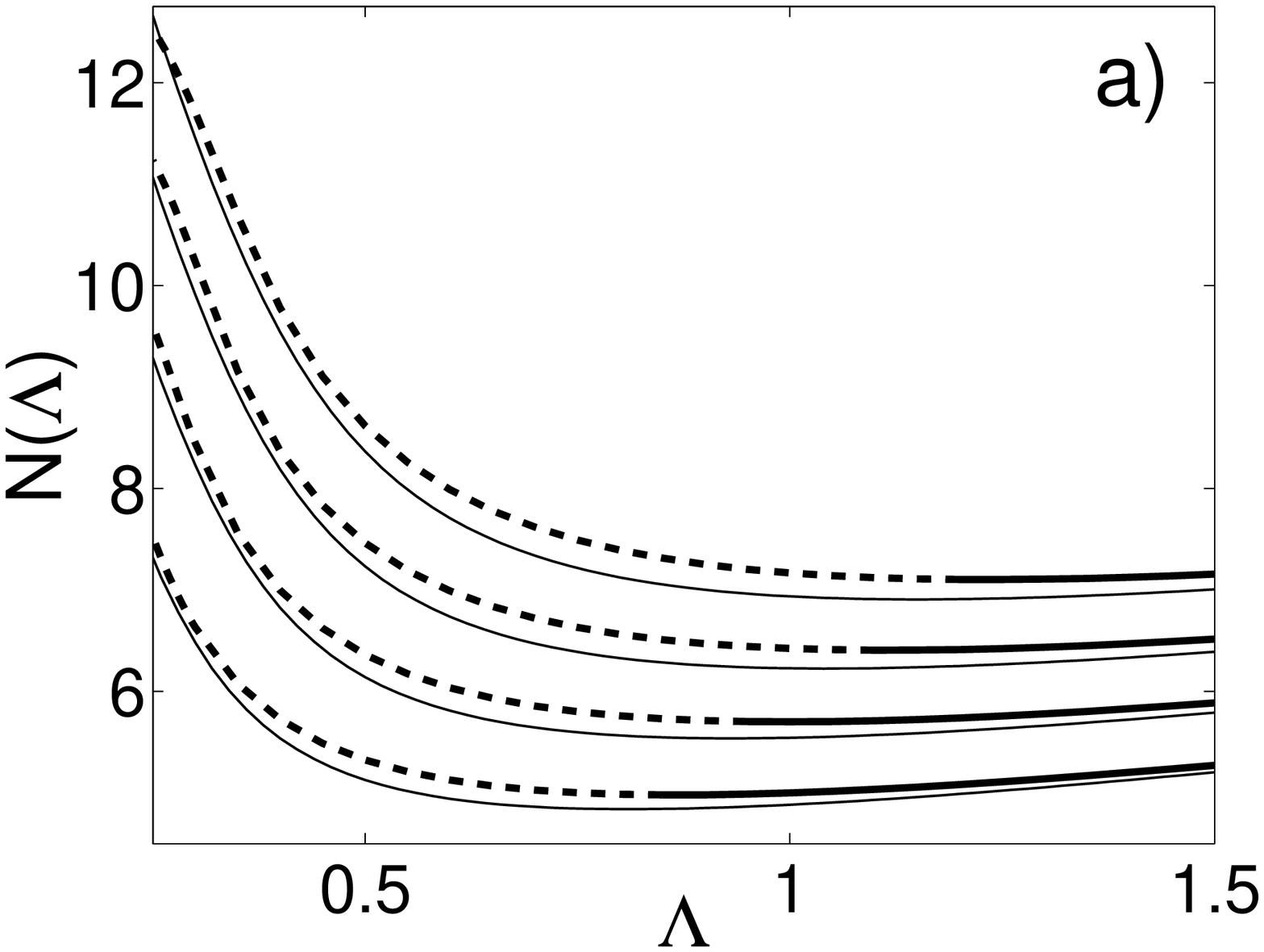}
}
\medskip
\centerline{
\includegraphics[width=6.cm,height=4cm,angle=0,clip]{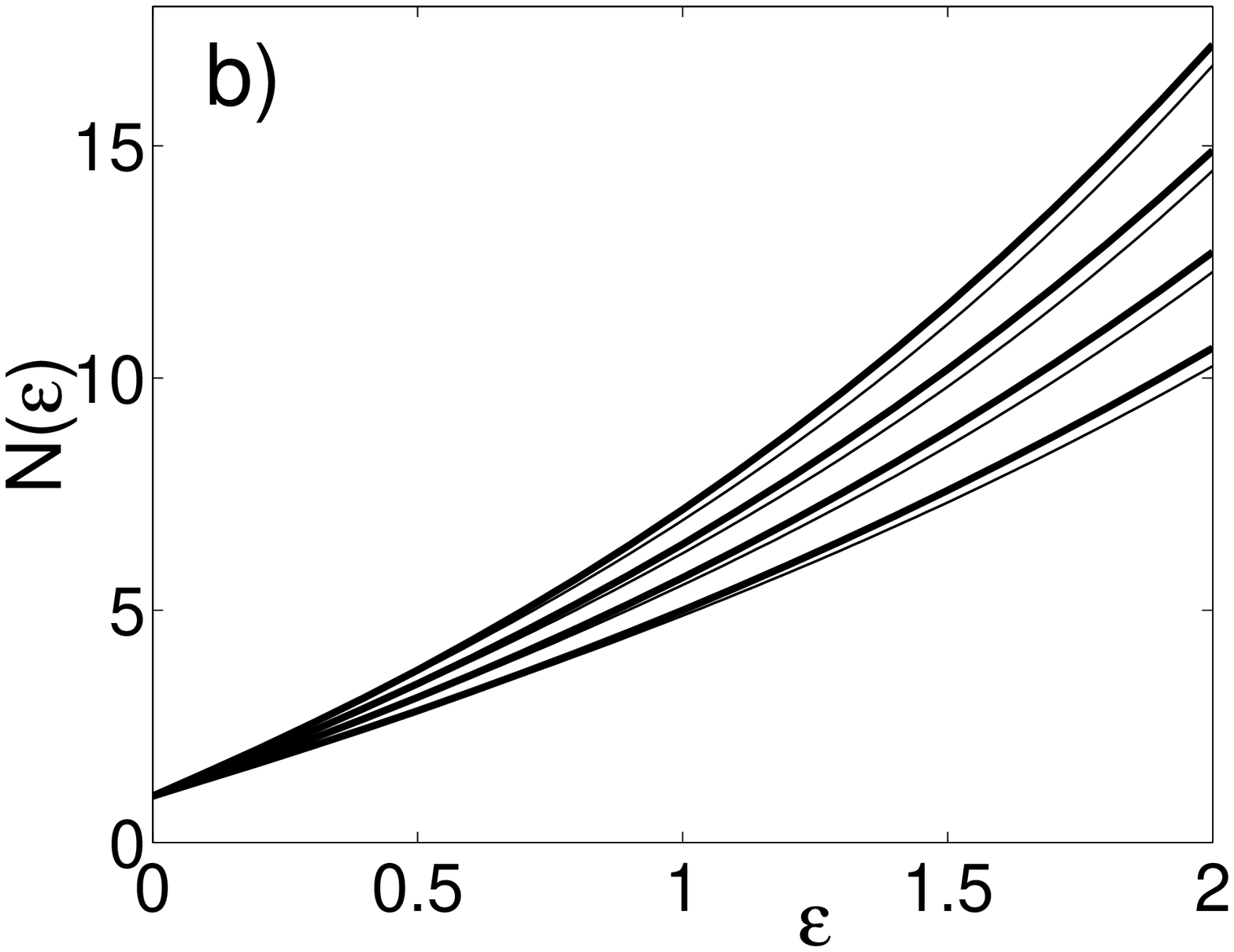}
}
\caption{a) Norm of the solution vs.\ $\Lambda$
for several values of the anisotropy and fixed coupling
strength $\epsilon=1$. For all the panels
in this figure the anisotropy values are
$\alpha=1.5,1.25,1,0.75$,
respectively, for each curve from top to bottom.
Thick lines (solid and dashed) represent
direct numerical results and the thin lines represent the VA.
The dashed lines correspond to unstable soliton solutions
%
Note that the sign of the slope of $N(\Lambda)$ reflects the
stability of the soliton solutions as predicted by the VK criterion.
b) The norm of the soliton solution as a function of the
coupling strength for fixed $\Lambda=1$.
Once again, thick lines represent direct numerical
results and thin lines illustrate the VA.
}
\label{NvsLambda_fig}
\end{figure}

Lastly, inequalities (\ref{cond}) and (\ref{cond2}) imply that the
above solutions indeed pertain to the strongly anisotropic model,
as the corresponding parameter $\alpha $ is large in the former
case, and small in the latter one. We also notice that the
condition (\ref{cond}) in the case of large $\alpha $, or its
counterpart (\ref{cond2}) in the opposite case of small $\alpha $,
is incompatible with the respective condition (\ref{condLambda}),
i.e. (as one would expect), the existence regions of the unstable
quasi-continuum solitons and stable strongly anisotropic ones have
no overlap.

For general $a$ and $b$, the variational equations (\ref{general}),
with the effective Lagrangian (\ref{Leff}), cannot be solved
explicitly and one has to find $(N,a,b)$ solutions numerically
for each $(\epsilon,\Lambda)$ pair.
In Fig.\ \ref{NvsLambda_fig} we compare the results
obtained from the VA with solutions obtained through numerically
solving the stationary equation (\ref{stationary}).
Fig.\ \ref{NvsLambda_fig}.a depicts the norm of the soliton solutions
as a function of the propagation constant $\Lambda$ for several
values of the anisotropy parameter $\alpha$ and for
constant coupling ($\epsilon=1$). As may be
noticed from the figure, the VA (thin lines) provides a good
approximation to the actual solution (thick lines).
We also checked the stability of the constructed solutions
by following the largest real eigenvalue (see also details
below).
of the linearized problem defined in Eq.\ (\ref{lambda}).
Stable solutions are depicted with solid lines while
unstable solutions correspond to dashed lines.
As is clear from the figure, the slope of $N(\Lambda)$
predicts the stability of the solution according to the VK criterion
(see above). Furthermore, since the VA gives a good
approximation of $N(\Lambda)$, it is possible to obtain
a good estimate for the transition from stable to unstable
solutions, as $\Lambda$ is decreased, using the VA together
with the VK criterion.
Finally, in Fig.\ \ref{NvsLambda_fig}.b we fix $\Lambda=1$ and
perform a similar calculation by varying the coupling strength
$\epsilon$. Again, the VA (thin lines) approximates remarkably
well the norm of the solutions (thick lines).

\section{Fundamental solitons and vortex crosses: numerical results}

We start numerical computations with a single excited site in the
AC limit, and continue the solution in $\epsilon $ (for a fixed
value of the anisotropy parameter $\alpha $). The objective is to
construct regular site-centered discrete solitons, with the
anticipation that, as is known for the isotropic model ($\alpha
=1$), the solitons will be stable up to a critical value of the
coupling constant, i.e., at $\epsilon <\epsilon _{\mathrm{cr}}$
\cite{MIW2,2d}. At $\epsilon >\epsilon _{\mathrm{cr}}$, the discrete
solitons are found to be unstable due to a real eigenvalue arising
in the linearization around the soliton. In the numerical part of
the work (unlike the VA considered above), we fix $\Lambda =1$ in
Eq.\ (\ref{stateqn}), using the scaling invariance of Eq.\ (\ref{deq1}),
and examine how $\epsilon _{\mathrm{cr}}$ is
affected by the variation of $\alpha $. The results will be
summarized in the form of two-parameter diagrams that chart
regions of stable and unstable discrete states.

\begin{figure}[t]
\centerline{
\includegraphics[width=6.cm,height=5cm,angle=0,clip]{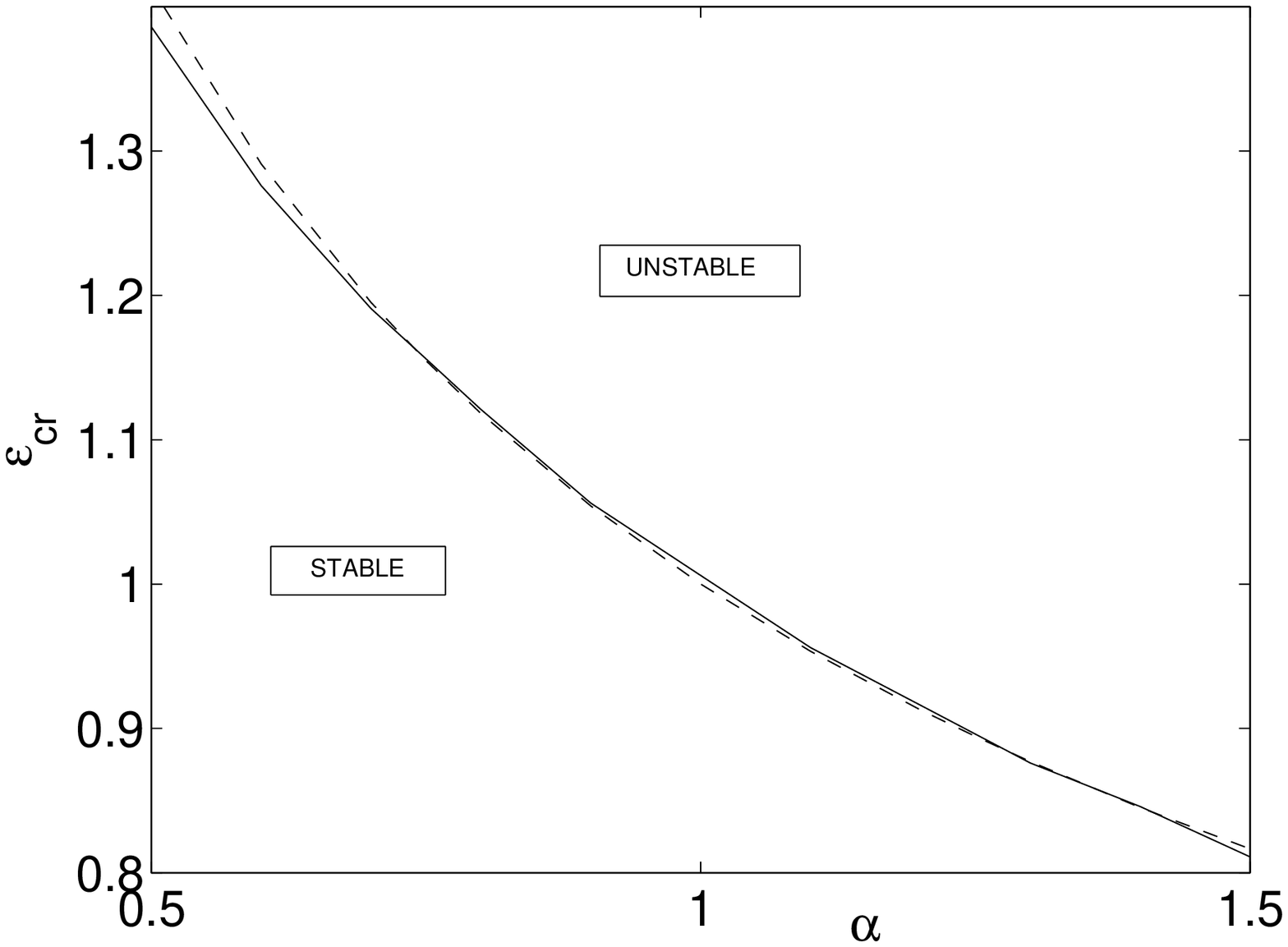}
}
\medskip
\centerline{
\includegraphics[width=5.8cm,height=4cm,angle=0,clip]{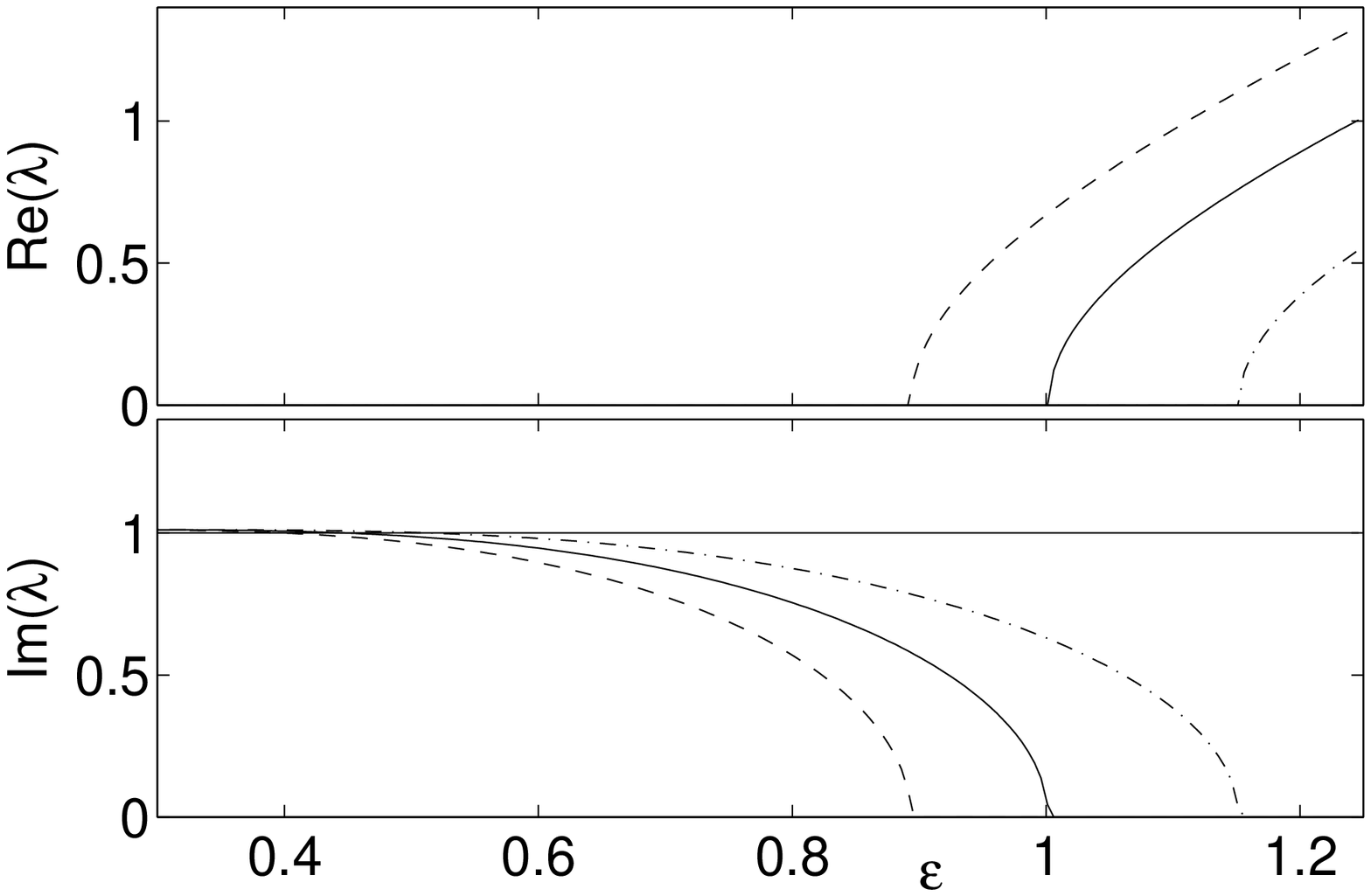}
}
\medskip
\centerline{
~~\includegraphics[width=6cm,height=5.25cm,angle=0,clip]{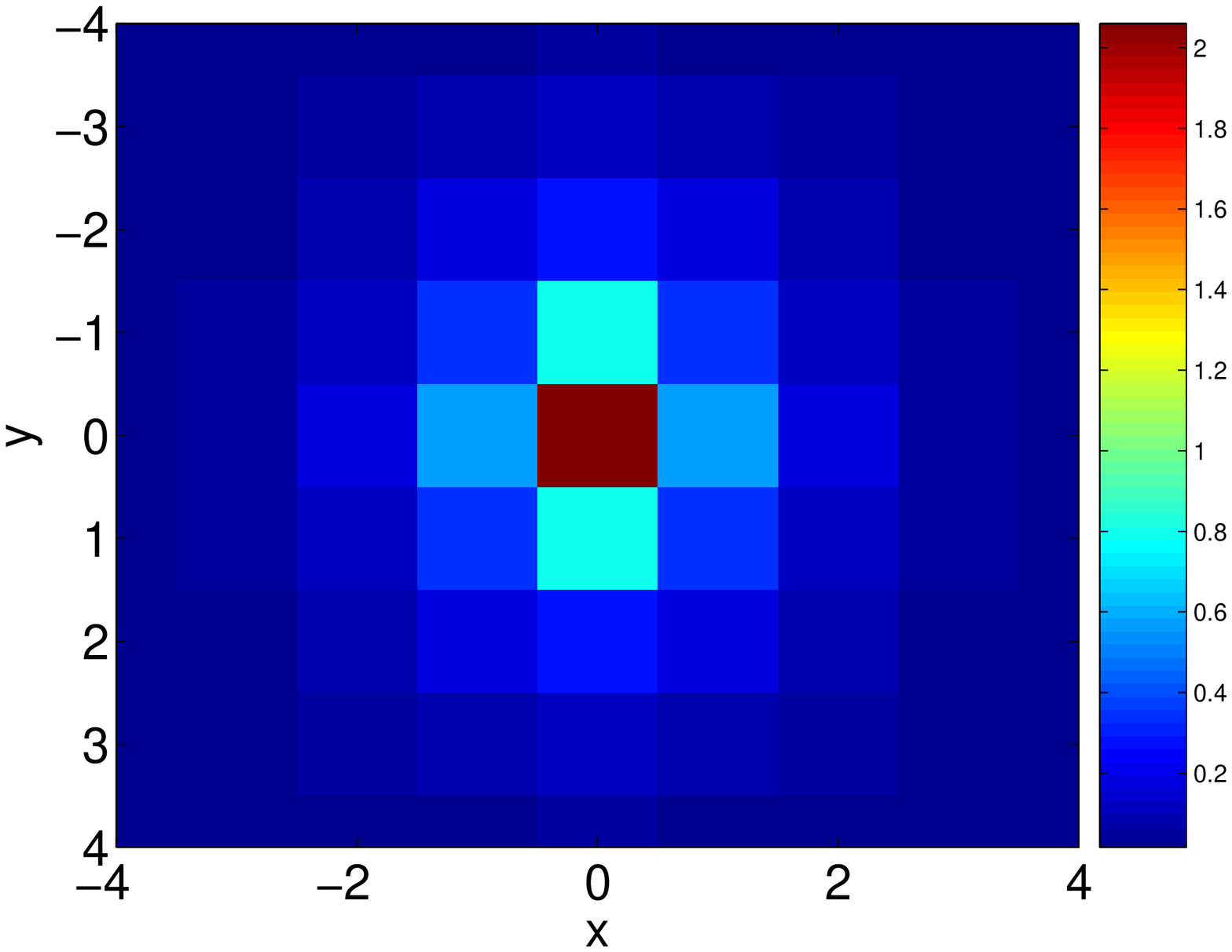}
}
\caption{The soliton line in the top panel shows the critical
value of $\protect\epsilon $ (the border between stable and
unstable discrete solitons) as a function of $\protect\alpha $;
the dashed line is $\protect\epsilon
=1/\protect\sqrt{\protect\alpha }$. The middle panel shows
how the real and imaginary parts of the stability eigenvalue,
$\protect\lambda _{r}$ and $\protect\lambda _{i}$, depend on
$\protect\epsilon $ for $\protect\alpha =1.25,~1$, and $0.75$
(dashed, solid, and dash-dotted curves, respectively). The
bottom panel shows an example of the discrete soliton found for
$\protect\epsilon =1$ and $\protect\alpha =1.5$.} \label{dfig1}
\end{figure}

For regular discrete solitons, such a diagram is presented in
Fig.\ \ref{dfig1}. The top panel illustrates the fact that the increase
of $\alpha $ gradually destabilizes the solitons, i.e., $\epsilon
_{\mathrm{cr}}$ decreases with increasing $\alpha $.
Interestingly, the respective dependence is very well approximated
by an empirical relation $\epsilon _{\mathrm{cr}}=1/\sqrt{\alpha }$.
More accurately, the best fit to this numerical dependence is
given by: $\epsilon _{\mathrm{cr}}\approx 0.999\alpha ^{-0.488}$.
The middle panel in Fig.\ \ref{dfig1} illustrates in more
detail some special cases of this dependence for $\alpha =1$
(solid lines), $\alpha =1.25$ (dashed lines) and $\alpha =0.75$
(dash-dotted lines).

We note that, in terms of the general equation (\ref{deq1}), the
cases of $\alpha <1$ and $\alpha >1$ are tantamount to each other,
as one may divide the equation by $\alpha $, mutually rename the
vertical and horizontal indices ($n$ and $m$), and then rescale
the equation to the form with $\alpha $ replaced by $1/\alpha $.
However, this transformation is not possible once we fix $\Lambda
\equiv 1$, which is why we report results below for both $\alpha
>1$ and $\alpha <1$.

For $\alpha =1$, an eigenvalue bifurcates from the edge of the
continuous spectrum at $\epsilon \approx 0.445$, and with further
increase of $\epsilon $ it moves towards the origin of the
spectral plane $(\lambda _{r},\lambda _{i})$ (the subscripts
denote the real and imaginary part of the eigenvalue) . It becomes
unstable, reaching the origin at $\epsilon \approx 1.006$. For $\alpha
=1.25 $, the first bifurcation occurs at $\epsilon \approx 0.398,$ and
the instability sets in at $\epsilon \approx 0.896$, whereas for $\alpha
=0.75$ the respective critical points (the appearance of the
eigenvalue, and its passage into the instability region) are found
at $\epsilon \approx 0.511$ and $1.156$, respectively. Notice that these
results are quite natural since, as $\alpha \rightarrow 0$, the
system becomes nearly one-dimensional, hence we expect the
destabilization point to approach its 1D counterpart. Thus, as the
1D discrete solitons are well-known to be stable up to the
continuum limit, one may expect that $\epsilon
_{\mathrm{cr}}\rightarrow \infty $ for $\alpha \rightarrow 0$. The
bottom panel of Fig.\ \ref{dfig1} shows an example of a
discrete soliton for $\alpha =1.5$ and $\epsilon =1$. Although the
anisotropy is hardly observed in this case, it can be traced
nevertheless; in particular, $u_{1,0}=0.785$, and $u_{0,1}=0.579$.

\begin{figure}[t]
\centerline{
\includegraphics[width=5.6cm,height=4.5cm,angle=0,clip]{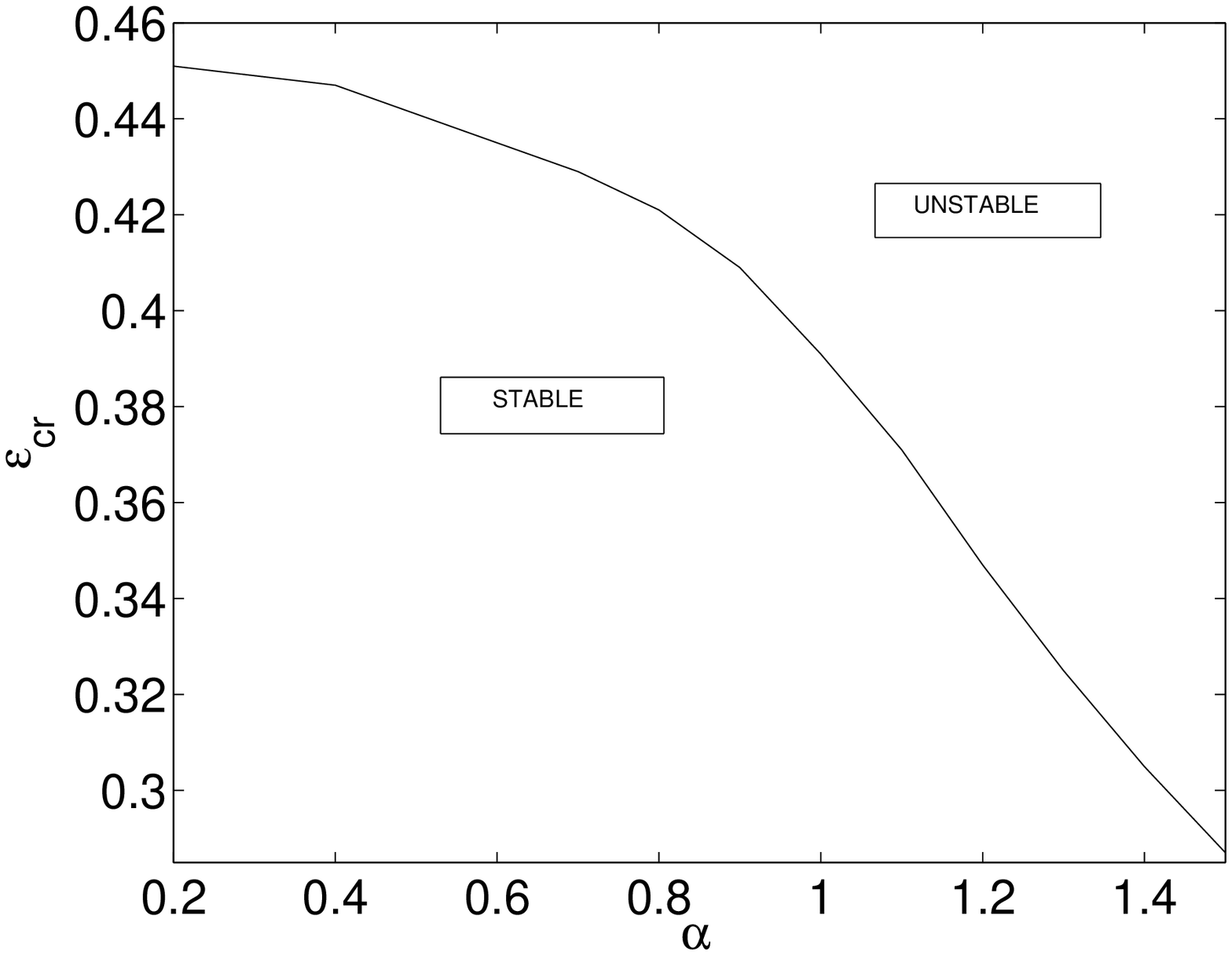}
}
\medskip
\centerline{
\includegraphics[width=5.8cm,height=4cm,angle=0,clip]{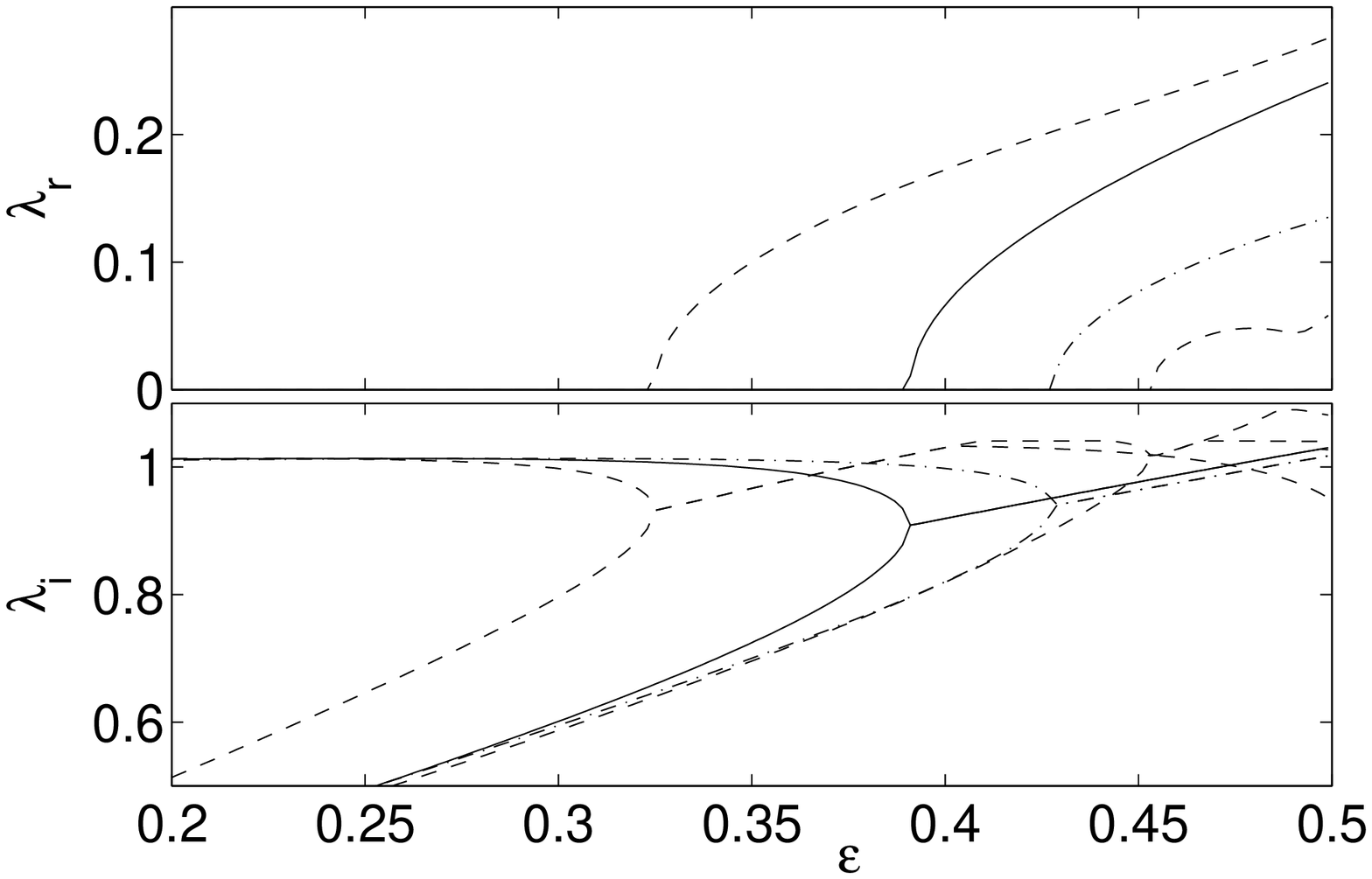}
}
\medskip
\centerline{
~~\includegraphics[width=6cm,height=5.25cm,angle=0,clip]{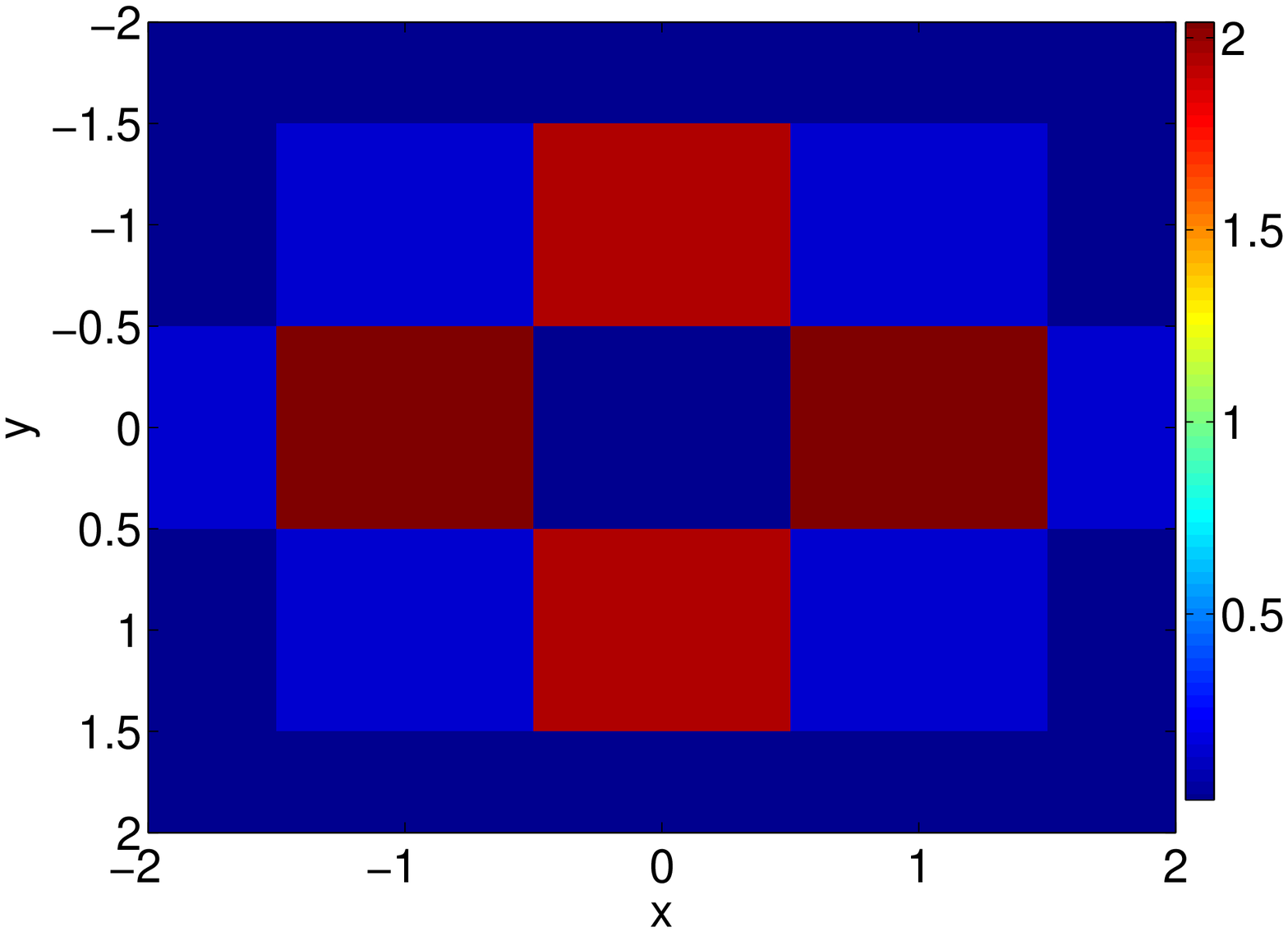}
}
\caption{
The top panel shows the critical value of
$\protect\epsilon $ separating the stable and unstable discrete
vortices (on-site-centered ones, alias \textit{vortex crosses})
with $S=1$ as a function of $\protect\alpha $. The middle
panel shows how the real and imaginary parts of the eigenvalue
leading to the instability depend on $\protect\epsilon $ for
$\protect\alpha =1.3,~1$, and $0.7$ (dashed, solid and dash-dotted
curves, respectively). Notice that, for $\protect\alpha =1.3$,
there is a secondary instability arising for $\protect\epsilon
>0.433$. The bottom panel shows the
squared-absolute-value profile of the discrete vortex for
$\protect\epsilon =0.5$ and $\protect\alpha =0.2$.} \label{dfig2}
\end{figure}

Similar results can be obtained for on-site vortices (discrete
vortex solitons) with the topological charge $S=1$. In this
section, we consider the solitons in the form of the so-called
``vortex cross", with $u_{1,0}=1$, $u_{0,1}=\exp (i\pi /2)\equiv
i$, $u_{-1,0}=\exp (i\pi )\equiv -1$, $u_{0,-1}=\exp (i3\pi
/2)\equiv -i$ (and $u_{0,0}=0$, at the central point), excited in
the AC limit \cite{vort1}. There are interesting variations to
this problem, in comparison with the fundamental soliton. In
particular, the respective instability mechanism is different, as
it is caused by an eigenvalue bifurcating from the origin in the
spectral plane for $\epsilon \neq 0$, and eventually (upon
parametric continuation) colliding with the edge of the continuous
spectrum (or an eigenvalue bifurcating from the continuous
spectrum). The collision gives rise to a \emph{quartet} of
eigenvalues, through the so-called Hamiltonian-Hopf bifurcation
\cite{hamilton}. In the isotropic case ($\alpha =1$), it is known
that this instability sets in at $\epsilon _{\mathrm{cr}}\approx 0.39$
\cite{vort1}, while, in the present anisotropic model, we have
found that $\epsilon _{\mathrm{cr}} \approx 0.325$ for $\alpha =1.3$ and
$\epsilon _{\mathrm{cr}} \approx 0.429$ for $\alpha =0.7$. The respective
two-parameter diagram $(\epsilon _{\mathrm{cr}},a)$ is shown in
the top panel of Fig.\ \ref{dfig2}. The cases of $\alpha =1.3$,
$1$, and $0.7$ (dashed, solid and dash-dotted, dashed-dotted
curves, respectively) are shown in the middle panel.
The bottom panel of the figure illustrates the
squared-amplitude profile of the discrete vortex for $\alpha =0.2$
and $\epsilon =0.5$. The sites $(1,0)$ and $(0,1)$ have the
squared amplitudes $|u_{1,0}|^{2}=1.934$ and
$|u_{0,1}|^{2}=2.057$, respectively. Notice also that as $\alpha
\rightarrow 0$, a quasi-1D situation is again approached, where
the so-called twisted-localized mode (TLM) \cite{TLM}
configuration (alias an odd soliton) is a counterpart of the 2D
vortex. As one would expect, the critical point of the instability
departs from the value $\epsilon
_{\mathrm{cr}}^{(\mathrm{2D})} \approx 0.39$, corresponding to the
isotropic 2D case, towards the value corresponding to the
stability border of the 1D TLM solitons, which is $\epsilon
_{\mathrm{cr}}^{(\mathrm{1D})}\approx 0.433$.

\section{Fundamental vortex squares}

For the discrete solitons examined so far,
the difference between the isotropic and non-isotropic cases has
not been particularly dramatic; the anisotropy chiefly entailed a
smooth deformation of the instability-onset scenarios known for
the isotropic case. Therefore, the dynamical evolution triggered
by the instability is naturally expected to be similar to that in
previously studied isotropic cases \cite{solit,vort1,vort2,2d}.

Now we will give an example where the instability scenario and
dynamics are \textit{very different} from their isotropic
counterparts. We focus, in particular, on the off site-centered
vortex (alias ``vortex square") \cite{vort1,vort2}. The
vortex-square contours are characterized by their size $M$, which
is the number of lattice bonds that each side of the square
contour contains in the AC-limit pattern, from which the solution
family stems. Hence, the vortex square based, in the AC limit, on
the set of sites $(0,0),(1,0),(1,1),(0,1)$ is the $M=1$ contour.
The configuration with $S=1$ is written on this set by lending the
four sites the phases $0$, $\pi /2$, $\pi $ and $3\pi /2$,
respectively. The persistence of such configurations, as was
discussed in detail in Ref.\ \cite{dep}, is determined by whether
secular conditions (obtained from the Lyapunov-Schmidt theory
\cite{golub}), excluding the projection of eigenvectors in the
kernel of the linearization at $\epsilon =0$ to the solution at
finite $\epsilon $, are satisfied. In the isotropic case, to
leading order ($O$($\epsilon $)), these secular conditions are
found to be
\begin{equation}
0=f(\theta _{l})\equiv \sin (\theta _{l}-\theta _{l+1})+
\sin (\theta _{l}-\theta _{l-1})  \label{deq3}
\end{equation}
for $l=1,\dots ,N$ (with periodic boundary conditions), where $N=4M$ is the
number of sites participating in the contour and $\theta_l$
are their respective phases [cf.\ Eqs.\ (3.1)--(3.2) of
Ref.\ \cite{dep}].

One can then apply similar arguments to the present setting and derive
modified persistence criteria for the anisotropic model. For $M=1$, they are
\begin{equation}
0=f(\theta _{l})\equiv \left\{
\begin{array}{rcl}
\alpha \sin (\theta _{l}-\theta _{l+1})&+&\sin (\theta _{l}-\theta _{l-1}) \\[0.5ex]
l&=&2k+1,k=0,1 \\[2.0ex]
\sin (\theta _{l}-\theta _{l+1})&+&\alpha \sin (\theta _{l}-\theta_{l-1}) \\[0.5ex]
l&=&2k,k=1,2.\end{array}\right.   \label{deq3a}
\end{equation}
While Eqs.\ (\ref{deq3a}) may seem  a moderate modification of (\ref{deq3}),
there is a crucial (for stability purposes) difference. Indeed,
consider the linearization around the $S=1$ solution according to
Eq.\ (\ref{lambda}). It was proved in Ref.\ \cite{dep} that the
Jacobian matrix of the reduced set of Eqs.\ (\ref{deq3a}), defined
through $J_{lk}=\partial f_{l}/\partial \theta _{k}$, determines
leading-order corrections to $N-1$ eigenvalue pairs bifurcating
from the origin [one pair stays at the origin due to the
invariance of Eq.\ (\ref{deq1}) with respect to the phase shift],
since these eigenvalues satisfy the equation:
\begin{equation}
\lambda _{l}^{2}=2\epsilon \mu _{l},  \label{deq3b}
\end{equation}with $\mu _{l}$ the corresponding eigenvalues of the reduced $N\times N$
Jacobian $J_{lk}$. It is further easy to check that, for the
vortex square with $S=1$ and $M=1$, the \emph{entire} Jacobian
matrix consists of \emph{zeros}. More generally, as shown in Ref.\ \cite{dep},
this is the case for the square vortices of size $M$
with charge $S=M$, which for that reason were termed
``super-symmetric" vortices. Obviously, to determine the stability
of the vortices in this special case, one needs to go to
higher-order expansions. Typically, second-order reductions will
yield a non-trivial result for the stability of such
super-symmetric configurations, leading to eigenvalue dependences
$\lambda _{l}\propto \epsilon $ [rather than $\lambda _{j}\propto
\sqrt{\epsilon }$, as dictated by Eq.\ (\ref{deq3b}) in the generic
case].

The key variation to this theme stemming from the presence of the
anisotropy is that the matrix $J_{lk}$ has generically
non-vanishing elements in the \emph{lowest approximation} for
$\alpha \neq 1$; in other words, the isotropic lattice is a
\emph{degenerate} one for the supersymmetric solitons, and
arbitrarily weak anisotropy \emph{lifts this degeneracy}. As a
result, the eigenvalue bifurcations occur, typically, at the
leading order, rather than at the second-order perturbation
expansions, which was the case in the isotropic model. More
strikingly, considering a specific example, such as for $\alpha
=1.05$ (a very weak deviation from the isotropic case), we find
that the relevant angles (in radians) satisfying the conditions
(\ref{deq3a}) are $\theta _{1}=-0.0229$, $\theta _{2}=1.8577$,
$\theta _{3}=3.4285 $, and $\theta _{4}=4.6895$; the corresponding
$4\times 4$ Jacobian has two zero eigenvalues (one of which will
split to order $O(\epsilon )$, see below) and two nonzero ones,
$\pm 0.6403$. From the existence of the positive eigenvalue and
from Eq.\ (\ref{deq3b}), it immediately follows that the $S=M=1$
configuration is \textit{immediately unstable} (for all values of
$\epsilon $). This is in {\it complete} contrast with the super-symmetric
vortex in the isotropic model, which has two imaginary eigenvalue
pairs (bifurcating at the second-order reduction), $\lambda
\approx \pm 2i\epsilon $, and is \textit{linearly stable }for
$\epsilon <\epsilon _{c}\approx 0.38$.

\begin{figure}[t]
\includegraphics[width=6.cm,height=5cm,angle=0,clip]{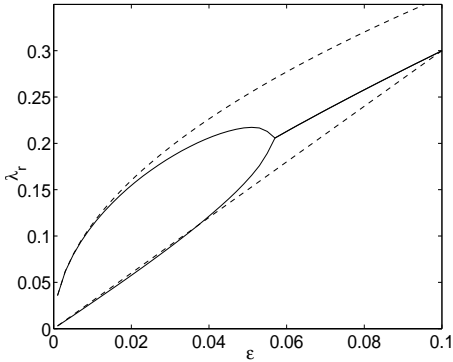}
\caption{For the $S=1$ super-symmetric square vortex (one with the
size $M=1$), two real stability eigenvalues are shown as functions
of $\protect\epsilon $ for $\protect\alpha =1.05$. The numerical
and analytical results (see text) are displayed, respectively, by
the solid lines and dashed lines.} \label{dfig2off}
\end{figure}

From here, we conclude that the anisotropy can play a critical
role in destabilizing configurations that would be very robust
ones in the isotropic limit. Furthermore, this can happen
arbitrarily close to the isotropic limit (that turns out to be a
very delicate one), given the nature of the argument presented
above. We also note in passing that in the anisotropic case
examined above, there is yet another real eigenvalue pair which is
$\lambda \approx \pm 3\epsilon $ for small $\epsilon $ (this pair
stems from the higher-order reduction, in agreement with the
prediction of the reduced Jacobian). These two eigenvalue pairs
eventually collide at $\epsilon =0.057$, resulting in a
Hamiltonian Hopf bifurcation to an eigenvalue quartet which is
present in the stability spectrum at $\epsilon >0.057$. This
phenomenology is shown in Fig.\ \ref{dfig2off}. The leading-order
prediction for the most unstable eigenvalue is in good agreement
with the full numerical result for small values of $\epsilon $.
For higher values of $\epsilon $, the second-order corrections
that we do not examine here in detail come into play and lead to
the Hamiltonian Hopf bifurcation.

To directly compare the dynamics between the isotropic and weakly
anisotropic (yet unstable) case for the super-symmetric vortex, we have
performed numerical simulations. Detailed simulations are reported in this
work only for the super-symmetric cases (see also the next section), since
for all other states anisotropy operates as a regular perturbation, see
above; as a result, instabilities of the other states may be shifted due to
the anisotropy, but structurally the phenomenology remains the same.

\begin{figure}[t]
\includegraphics[width=6.cm,height=6cm,angle=0,clip]{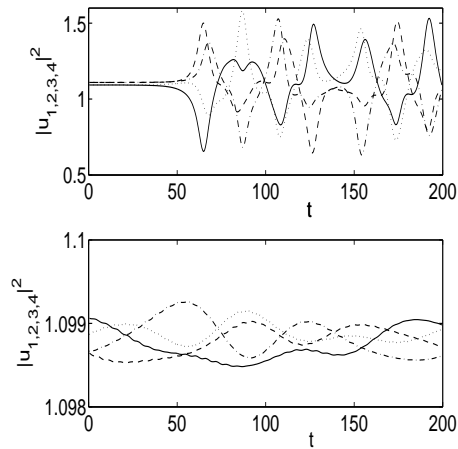}
\caption{The dynamics of an initially very weakly perturbed super-symmetric
vortex with $S=M=1$, principally based on four lattice sites that form an
elementary cell (the sites are labeled as $1,2,3,4$). The time evolution of
the squared absolute value of the fields at these sites is shown in the top
panel for a weakly anisotropic model, with $\protect\alpha =1.05$, and for
its isotropic counterpart ($\protect\alpha =1$) in the bottom panel. In both
cases, the same uniformly distributed, random initial perturbation of
amplitude $10^{-4}$ was added to the solution at $t=0$ to excite possible
instabilities. Clearly, the vortex on the weakly anisotropic lattice becomes
unstable at $t>50$, while in the isotropic case the perturbation remains
bounded and small at all times. In these examples, the intersite lattice
coupling constant is $\protect\epsilon =0.025$.}
\label{dfig2off2}
\end{figure}

For the delicate super-symmetric vortex square, the dynamics
altered by the anisotropy is indeed found to be dramatically
different from the isotropic case. This is illustrated by
Fig.\ \ref{dfig2off2}, for the vortex square with $S=M=1$, carried (in
the AC limit) by $4$ sites. The time dynamics of the squared
absolute value of the field at the main sites is shown in the
figure for a weakly anisotropic model, with $\alpha =1.05$, and
its isotropic counterpart (top and bottom panels, respectively).
Stark contrast between the instability developing for $t>50$ in
the former case, versus the complete stability for all times in
the latter (isotropic) system, is obvious (notice the difference
in the scales of vertical axes between the two panels). In the
linear approximation, these results are well predicted by the
above theory.

\section{Higher-order vortices}

\begin{figure}[t]
\includegraphics[width=6.cm,height=5cm,angle=0,clip]{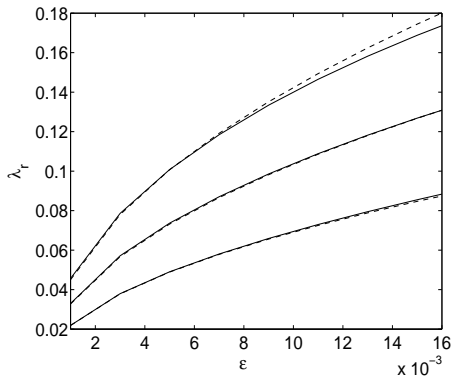}
\caption{For the $S=M=2$ supersymmetric vortex, the three real
eigenvalues are displayed as functions of $\protect\epsilon $ for
$\protect\alpha =1.05$. The solid and dashed lines depict the
numerical and analytical results.} \label{dfig3}
\end{figure}

We now give a summary of results for vortices with higher values
of the topological charge. First, we consider the $S=M=2$ super-symmetric
vortex populating the sites $(1,0)$, $(1,1)$, $(0,1)$, $(-1,1)$,
$(-1,0)$, $(-1,-1)$, $(0,-1)$, $(1,-1)$ in the AC limit, with a
phase shift of $\pi /2$ between adjacent sites (in the isotropic
model). The latter provides for a total phase gain of $4\pi $
around a closed path surrounding the origin. This type of the
configuration with $S=M=2$ was identified in Ref.\ \cite{dep} as
possessing a real eigenvalue pair with $\lambda
_{r}=\pm \sqrt{\sqrt{80}-8}\epsilon $, in excellent agreement with
numerical computations. However, the presence of the small
anisotropy for $\alpha \neq 1$ again strongly affects the vortex
for reasons similar to the ones presented above. In this case, the
reductions leading to the perturbed dynamics in the anisotropic
model are described by the following persistence conditions: \
\begin{equation}
0=f(\theta _{l})\equiv \left\{
\begin{array}{rcl}
\sin (\theta _{l}-\theta _{l+1})&+&\sin (\theta _{l}-\theta _{l-1})\\[0.5ex]
i&=&2k+1,k=0,1,2,3, \\[2.0ex]
\alpha \sin (\theta _{l}-\theta _{l+1})&+&\sin (\theta _{l}-\theta _{l-1}) \\[0.5ex]
i&=&4k+2,k=0,1, \\[2.0ex]
\sin (\theta _{l}-\theta _{l+1})&+&\alpha \sin (\theta _{l}-\theta_{l-1}) \\[0.5ex]
i&=&4k+4,k=0,1,\end{array}\right.   \label{deq10}
\end{equation}
cf.\ Eqs.\ (\ref{deq3a}). In this expression $\theta _{l}$ is the phase of
the field at each of the eight above-mentioned sites (where, in the order
the sites were mentioned, the corresponding index is $l=1,2,\dots ,8$).
Furthermore, as discussed above, the analysis performed in Ref.\ \cite{dep}
can be used to show that the linear stability eigenvalues for such a vortex soliton
will be given, to the leading order, by Eq.\ (\ref{deq3b}).
Using this prediction, even in the weakly anisotropic case (e.g.,
for $\alpha =1.05$) one finds that the corresponding $8\times 8$
Jacobian possesses three real $O(\sqrt{\epsilon })$ eigenvalues,
which result in an instability (contrary to the single real
$O(\epsilon )$ eigenvalue in the $\alpha =1$ case). Hence, once
again, the anisotropy results in a significant destabilization of
the super-symmetric vortex, in comparison to the isotropic model.
As a specific example, we show in Fig.\ \ref{dfig3}) the situation
for $\alpha =1.05$. The solution of Eqs.\ (\ref{deq10}) yields
$\theta _{1}=0.218$, $\theta _{2}=1.967$, $\theta _{3}=3.182$,
$\theta _{4}=4.397$, $\theta _{5}=6.145$, $\theta _{6}=7.894$,
$\theta _{7}=9.109$, $\theta _{8}=11.036$, which, in turn, results
in a Jacobian with the 3 real eigenvalues
$\mu=
\{1.0145,0.5357,0.2391\}$. The comparison of the
numerical prediction for the eigenvalue dependence on $\epsilon $
versus the corresponding analytical prediction (solid and dashed
lines, respectively) based on the above results is given in
Fig.\ \ref{dfig3}, demonstrating a very good agreement between the two.

\begin{figure}[t]
\includegraphics[width=6.cm,height=6cm,angle=0,clip]{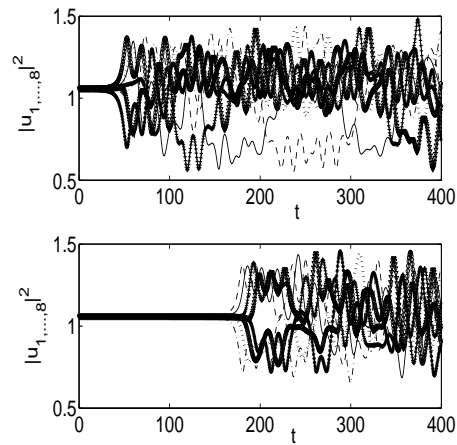}
\caption{Same as in Fig.\ \protect\ref{dfig2off2} but for the supersymmetric
vortex of the $S=M=2$ type. The four lines (solid, dashed, dotted, and
dash-dotted) and four symbols (circles, pluses, stars and triangles) are
used to denote the squared absolute values of the field at the eight sites
carrying the vortex in the anisotropic (top) model and its isotropic
counterpart (bottom) for $\protect\epsilon =0.015$.}
\label{dfig4}
\end{figure}

To highlight the substantial differences between the dynamics in
the isotropic and anisotropic models, we have performed numerical
simulations of the super-symmetric vortex with $S=M=2$. In this
case, the evolution of the field at the eight basic sites is shown
in the top panel of Fig.\ \ref{dfig4} for $\alpha =1.05$, and in
the bottom panel for $\alpha =1$. In the former case, for the
coupling strength $\epsilon =0.015$ considered here, the three
unstable eigenvalues for $\alpha =1.05$ are $\lambda =0.1688$,
$\lambda =0.1258$ and $\lambda =0.0855$, while in the latter case
(isotropic model), the only unstable eigenvalue is a much smaller
one, $\lambda =0.0146$. Naturally, we observe the instability
setting in much earlier in the anisotropic model (at $t~\,_{\sim
}^{>}~30$) than in the isotropic one (at $t~\,_{\sim }^{>}~160$).

One may be wondering whether the strong dynamical effect of the weak
anisotropy should be attributed to the super-symmetry of the
vortex, or maybe just the specific type of contour which
carries the vortex. To check this, we have also considered the
vortex with $S=3$ sitting on the same $M=2$ contour. Given the
lack of the super-symmetry in the latter case, the bifurcation of
the relevant $7$ ($=N-1$) eigenvalue pairs occurs at the
leading-order reduction and all of them are proportional to $\pm
i\sqrt{\epsilon }$. More specifically, for the largest pair in the
isotropic case (for instance), the proportionality factor is
$2.3784$. In the anisotropic case with $\alpha =1.05$, the seven
pairs remain on the imaginary axis, being slightly perturbed due
to $\alpha \neq 1$. For instance, the largest one among them is
now $\lambda =\pm 2.3943\cdot i\sqrt{\epsilon }$. On the other
hand, for $\alpha =0.95$, the largest eigenvalue pair is $\lambda
=\pm 2.3647\cdot i\sqrt{\epsilon }$. This also is in line with our
above results on the fundamental discrete soliton and vortex
cross, since it indicates that, for $\alpha >1$, the collision of
this eigenvalue with the continuous spectrum (which leads to the
Hamiltonian Hopf bifurcation) will occur at smaller $\epsilon $,
the opposite being true for $\alpha <1$. Hence the stability
diagram of the $S=3,M=2$ vortex square is quite similar to that
shown for the fundamental soliton and vortex cross in
Figs.\ \ref{dfig1} and \ref{dfig2} (therefore, it is not shown here).

\section{Conclusions}

In this work, we have examined effects of anisotropy on lattice nonlinear
dynamical systems supporting discrete solitons and vortices. The
two-dimensional discrete nonlinear Schr\"{o}dinger equation was used as a
paradigm model. The variational approximation was developed for fundamental
solitons, showing (by means of the Vakhitov-Kolokolov criterion) that broad
quasi-continuum ones are unstable, while strongly anisotropic solitons
are stable. By means of numerical methods, we
have found that usual localized states, such as the fundamental discrete
solitons and vortex crosses, are only mildly affected by the anisotropy,
which results in a modified stability region (reduced when one direction
features a stronger coupling than the isotropic limit, and augmented when
the coupling along this direction is weaker). General phenomenology for such
states is similar to that for their counterparts on the isotropic lattice.

The main finding reported in the present work is that the assumption about
mild deformation of the stability region induced by  weak anisotropy
is not valid for the delicate super-symmetric vortex states residing on
square contours, in the case when the vorticity $S$ is equal to the
contour's size $M$. In this special case, the degeneracy of the
leading-order existence conditions (dictated by Lyapunov-Schmidt theory)
specific to the isotropic case is broken by the anisotropy. This, in turn,
results in a dramatically different behavior (as a function of the intersite
coupling constant) of the corresponding linear
stability eigenvalues, in terms of
both the order of their bifurcation, and the number of real eigenvalues. As
a consequence, the supersymmetric vortex-square structure that was
marginally stable in the isotropic case is found to be strongly unstable
even on the weakly anisotropic lattice. Similarly, the supersymmetric vortex
with $S=M=2$ is found to be much more unstable in the anisotropic case in
comparison to its isotropic counterpart.

The most natural systems for experimental observation of the results
predicted in this work are deep optical lattices trapping BECs, and bundled
sets of nonlinear optical waveguides (the latter have been recently created
experimentally \cite{Lederer}). Anisotropic lattices can also be induced
in photorefractive media, but this medium should be considered separately,
in view of the different (saturable) character of the optical nonlinearity
in this case. Such investigations are currently in progress and will be
reported elsewhere.

A further natural extension of this work would be to examine effects of
anisotropy in three-dimensional lattices on discrete solitons, vortices,
dipoles and quadrupoles of various types, octupoles, and more exotic
localized configurations, that were recently investigated for the isotropic
case in Refs.\ \cite{vortex3d}.

~

\section*{ACKNOWLEDGEMENTS}

The work of B.A.M.\ was supported in a part by the Israel Science Foundation
through the grant No. 8006/03.
P.G.K.\ gratefully acknowledges support
from NSF-DMS-0204585, NSF-CAREER. R.C.G.\ and P.G.K.\ also
acknowledge support from NSF-DMS-0505663. Work at Los Alamos is supported
by the US DoE.


\begin{thebibliography}{99}
\bibitem{reviews} S. Aubry, Physica D \textbf{103}, 201, (1997); S. Flach
and C. R. Willis, Phys. Rep. \textbf{295} 181 (1998); D. Hennig and G.
Tsironis, Phys. Rep. \textbf{307}, 333 (1999); P. G. Kevrekidis, K. {\O }.
Rasmussen, and A. R. Bishop, Int. J. Mod. Phys. B \textbf{15}, 2833 (2001).

\bibitem{reviews1} D. N. Christodoulides, F. Lederer and Y. Silberberg,
Nature \textbf{424}, 817 (2003); Yu. S. Kivshar and G. P. Agrawal,
\textit{Optical Solitons: From Fibers to Photonic Crystals},
Academic Press (San Diego, CA, 2003).

\bibitem{reviews2} P. G. Kevrekidis and D. J. Frantzeskakis, Mod. Phys.
Lett. B \textbf{18}, 173 (2004); V. V. Konotop and V. A. Brazhnyi,
Mod. Phys. Lett. B \textbf{18} 627, (2004); P. G. Kevrekidis, R.
Carretero-Gonz\'{a}lez, D. J. Frantzeskakis, and I. G. Kevrekidis,
Mod. Phys. Lett. B \textbf{18}, 1481 (2004).

\bibitem{sievers} M. Sato and A. J. Sievers, Nature \textbf{432}, 486
(2004); M. Sato, B. E. Hubbard, A. J. Sievers, B. Ilic and H. G. Craighead,
Europhys. Lett. \textbf{66}, 318 (2004); M. Sato, B. E. Hubbard, A. J.
Sievers, B. Ilic, D. A. Czaplewski, and H. G. Craighead, Phys. Rev. Lett.
\textbf{90}, 044102 (2003).

\bibitem{reviews3} M. Peyrard, Nonlinearity \textbf{17}, R1 (2004).

\bibitem{voglis} N. Voglis, Mon. Not. Roy. Astr. Soc. \textbf{344}, 575
(2003).

\bibitem{solit} N. K. Efremidis, {\it et al.},
S. Sears, D. N. Christodoulides, J. W. Fleischer, and M. Segev, Phys. Rev. E
\textbf{66}, 046602 (2002); A. A. Sukhorukov, Yu. S. Kivshar, H. S
Eisenberg, and Y. Silberberg, IEEE J. Quantum Electron. \textbf{39}, 31
(2003).

\bibitem{vort1} B. A. Malomed and P. G. Kevrekidis, Phys. Rev. E \textbf{64}, 026601 (2001).

\bibitem{vort2} J. Yang and Z. Musslimani, Opt. Lett. \textbf{23}, 2094
(2003), P. G. Kevrekidis, B. A. Malomed, Z. G. Chen, and D. J.
Frantzeskakis, Phys. Rev. E \textbf{70}, 056612 (2004).

\bibitem{esolit} J. W. Fleischer, T. Carmon, M. Segev, N. K. Efremidis, and
D. N Christodoulides, Phys. Rev. Lett. \textbf{90} 023902 (2003); H. Martin,
E. D. Eugenieva, Z. G. Chen, and D. N. Christodoulides, Phys. Rev. Lett.
\textbf{92} 123902 (2004); J. K Yang, I. Makasyuk, A. Bezryadina, and Z.
Chen, Opt. Lett. \textbf{29}, 1662 (2004); Z. G. Chen, H. Martin, E. D.
Eugenieva, J. J. Xu, and A. Bezryadina, Phys. Rev. Lett. \textbf{92} 143902
(2004), Z. G. Chen, A. Bezryadina, I. Makasyuk, and J. K. Yang, Opt. Lett.
\textbf{29} 1656 (2004); J. Yang, 
I. Makasyuk I, P. G. Kevrekidis, H. Martin, B. A. Malomed, D. J.
Frantzeskakis, and Z. G. Chen, Phys. Rev. Lett. \textbf{94}, 113902 (2005).

\bibitem{evort} D. N. Neshev, T. J. Alexander, E. A. Ostrovskaya, Yu. S.
Kivshar, H. Martin, I. Makasyuk, and Z. G. Chen, Phys. Rev. Lett.
\textbf{92}, 123903 (2004); J. W. Fleischer, G. Bartal, O. Cohen,
O. Manela, M. Segev, J. Hudock, and D. N. Christodoulides, Phys.
Rev. Lett. \textbf{92} (2004) 123904.

\bibitem{becd} S.\ Burger, K.\ Bongs, S.\ Dettmer, W.\ Ertmer, K.\
Sengstock, A.\ Sanpera, G. V.\ Shlyapnikov, and M.\ Lewenstein, Phys.\ Rev.\
Lett.\ \textbf{83}, 5198 (1999); J.\ Denschlag, J. E.\ Simsarian, D. L.\
Feder, C. W.\ Clark, L. A.\ Collins, J.\ Cubizolles, L.\ Deng, E. W.\
Hagley, K.\ Helmerson, W. P.\ Reinhardt, S. L.\ Rolston, B. I.\ Schneider,
and W. D.\ Phillips, Science \textbf{287}, 97 (2000); B. P.\ Anderson, P.
C.\ Haljan, C. A.\ Regal, D. L.\ Feder, L. A.\ Collins, C. W.\ Clark, and E.
A.\ Cornell, Phys.\ Rev.\ Lett.\ \textbf{86}, 2926 (2001).

\bibitem{becb} K. E.\ Strecker, G. B.\ Partridge, A. G.\ Truscott, and R.
G.\ Hulet, Nature \textbf{417}, 150 (2002); L.\ Khaykovich, F.\ Schreck, G.\
Ferrari, T.\ Bourdel, J.\ Cubizolles, L. D.\ Carr, Y.\ Castin, and C.\
Salomon, Science \textbf{296}, 1290 (2002).

\bibitem{becg} B.\ Eiermann, Th.\ Anker, M.\ Albiez, M.\ Taglieber, P.\
Treutlein, K.-P.\ Marzlin, and M. K.\ Oberthaler, Phys.\ Rev.\ Lett.\
\textbf{92}, 230401 (2004).

\bibitem{boris} B. B. Baizakov, B. A. Malomed, and M. Salerno, Europhys.
Lett. \textbf{63}, 642 (2003); B. B. Baizakov, B. A. Malomed and M. Salerno,
Phys. Rev. A \textbf{70}, 053613 (2004).

\bibitem{old2d} See, e.g., M. Greiner, I. Bloch, O. Mandel, T. W. Hansch, T.
Esslinger, Appl. Phys. B \textbf{73}, 769 (2001); M. Greiner, I. Bloch, O.
Mandel, T. W. Hansch, T. Esslinger, Phys. Rev. Lett. \textbf{87}, 160405
(2001).

\bibitem{mackay} R. S. MacKay and S. Aubry, Nonlinearity \textbf{7} (1994)
1623.

\bibitem{VK} M. G. Vakhitov and A. A. Kolokolov, Izv. Vuz. Radiofiz. \textbf{16} (1973) 1020
[in Russian; English translation: Sov. J. Radiophys. Quantum
Electr. \textbf{16} (1973) 783]; L. Berg{\'{e}}, Phys. Rep. \textbf{303},
260 (1998).

\bibitem{Bakhtiyor} B. B. Baizakov, B. A. Malomed, and M. Salerno, Europhys.
Lett. \textbf{63}, 642 (2003); Phys. Rev. A \textbf{70}, 053613 (2004).

\bibitem{Bakhtiyor2} B. B. Baizakov, M. Salerno, and B. A. Malomed, in
\textit{Nonlinear Waves: Classical and Quantum Aspects} , ed. by
F. Kh. Abdullaev and V. V. Konotop, p. 61 (Kluwer Academic
Publishers: Dordrecht, 2004); also available at
http://rsphy2.anu.edu.au/$\sim$asd124/\-Baizakov\_2004\_61\_NonlinearWaves.pdf.

\bibitem{MIW} B. A. Malomed and M. I. Weinstein, Phys. Lett. A \textbf{220},
91 (1996).

\bibitem{MIW2} M. I. Weinstein, Nonlinearity \textbf{12}, 673 (1999).

\bibitem{2d} S. Flach, K. Kladko and R.S. MacKay,
Phys. Rev. Lett. {\bf 78}, 1207 (1997).
P. G. Kevrekidis, K. {\O }. Rasmussen and A. R. Bishop, Phys.
Rev. E \textbf{61}, 2006 (2000); P. G. Kevrekidis, K. {\O }. Rasmussen, and
A. R. Bishop, Math. Comp. Simul. \textbf{55}, 449 (2001).

\bibitem{hamilton} J.-C. van der Meer, Nonlinearity \textbf{3}, 1041 (1990).

\bibitem{TLM} S. Darmanyan, A. Kobyakov and F. Lederer, Sov. Phys. JETP
\textbf{86}, 682-686 (1998); P. G. Kevrekidis, A. R. Bishop and K. {\O }.
Rasmussen, Phys. Rev. E \textbf{63}, 036603 (2001).



\bibitem{dep} D. E. Pelinovsky, P. G. Kevrekidis, and D. J. Frantzeskakis,
nlin.PS/0411016.

\bibitem{golub} M. Golubitsky and D. G. Schaeffer, \emph{Singularities and
Groups in Bifurcation Theory}, vol. 1, (Springer-Verlag, New York, 1985).

\bibitem{Lederer} T.\ Pertsch, U.\ Peschel, F.\ Lederer, J.\ Burghoff, M.\
Will, S.\ Nolte, and A.\ Tunnermann, Opt.\ Lett.\ \textbf{29}, 468 (2004).

\bibitem{vortex3d} 
P. G.\ Kevrekidis, B. A.\ Malomed, D. J.\ Frantzeskakis, and R.\
Carretero-Gonz\'{a}lez, Phys.\ Rev.\ Lett.\ \textbf{93}, 080403 (2004);
R.\ Carretero-Gonz\'{a}lez, P. G.\ Kevrekidis, B. A.\ Malomed, and D. J.
Frantzeskakis, Phys.\ Rev.\ Lett.\ \textbf{94}, 203901 (2005).

\end{thebibliography}
\end{document}